# The Electronic Structure and Band gap of LiFePO$_4$ and LiMnPO$_4$

Fei Zhou, Kisuk Kang, Thomas Maxisch, Gerbrand Ceder, Dane Morgan

Massachusetts Institute of Technology, Cambridge MA

**Abstract:**

Materials with the olivine Li$_x$MPO$_4$ structure form an important new class of materials for rechargeable Li batteries. There is significant interest in their electronic properties because of the importance of electronic conductivity in batteries for high rate applications. The density of states of Li$_x$MPO$_4$ (x = 0, 1 and M = Fe, Mn) has been determined with the *ab initio* GGA+U method, appropriate for these correlated electron systems. Computed results are compared with the optical gap of LiFePO$_4$, as measured using UV-Vis-NIR diffuse reflectance spectroscopy. The results obtained from experiment (3.8-4.0 eV) and GGA+U computations (3.7 eV) are in very good agreement. However, standard GGA, without the same level of treatment of electron correlation, is shown to make large errors in predicting the electronic structure. It is argued that olivines are likely to be polaronic conductors with extrinsically determined carrier levels and that their electronic conductivity is therefore not simply related to the band gap.

## 1. Introduction

Low cost, good stability, and competitive electrochemical properties make the olivine $Li_xMPO_4$ family an exciting new area for cathode development in Li rechargeable batteries.[1-6] However, a major challenge in using these materials seems to be their low electrical conductivity. For example, the most studied member of the olivine cathode family has been $Li_xFePO_4$, which in its pure form has very poor conductivity, greatly inhibiting high-rate applications.[7] Similar problems are believed to inhibit Li exchange from $Li_xMnPO_4$.[6,8] Efforts to increase conductivity of electrodes made from these materials have focused on particle size reduction,[3] intimate carbon coating,[9] and cation doping.[7,10] Significant disagreement exists on the origin of the low electronic conductivity. A*b initio* studies focusing on the band gap and effective hole or electron mass have found a small gap, or no gap at the Fermi level, which seems to be in contradiction to experiment.[6,8,11-13] For example, Xu, *et al.*[13] found that $LiFePO_4$ is a semi-metal, which seems surprising, given the experimentally observed lack of electrical conductivity. However, there is significant evidence that the local density approximation (LDA) and generalized gradient approximation (GGA), used in almost all previous studies on the electronic structure of these phosphates, cannot accurately reproduce their electronic structure, due to the very approximate treatment of the electron correlation in transition metal orbitals by LDA/GGA. In order to clarify the electronic structure of $LiFePO_4$ and $LiMnPO_4$, we will apply the more accurate GGA+U method to determine the density of states of these systems and compare to diffuse spectroscopy measurements of the $LiFePO_4$ band gap. The implications of possible polaronic electrical conductivity will also be discussed.

## 2. Computational Methods

All calculations shown are performed within the Generalized Gradient Approximation (GGA)[14] or GGA+U,[15,16] with the projector-augmented wave (PAW) method[17,18] as implemented in the Vienna Ab-initio Simulation Package.[19] An energy cut-off of 500 eV and appropriate *k*-point mesh were chosen so that the total ground state energy is converged to within 3 meV per formula unit. In order to understand the possible impact of magnetic structure, we have performed calculations with both ferromagnetic (FM) and antiferromagnetic (AFM) orderings. The AFM ordering for all calculations was taken from the magnetic states of LiFePO$_4$ and LiMnPO$_4$, which have been determined experimentally to be AFM within the approximately simple square lattices formed by the planes of transition metal cations in the olivine structure.[20,21] For completeness we note that the low-temperature magnetic state of FePO$_4$ is non-collinear and slightly different from LiFePO$_4$[21], and that at higher temperatures all these systems will have magnetic disorder. For clarity, all density of states plots are presented in the FM ordering. All the atoms and cell parameters of each structure are fully relaxed, and the lattice parameters for the Li$_x$MPO$_4$ compounds, M = Fe and Mn, x = 0 and 1, both GGA and GG+U, are given in reference 22. The rotationally invariant[16] form of GGA+U is used with a spherically averaged double counting term.[23] These choices have previously been found to be effective in modeling phase stability and intercalation potential in olivine phosphates.[22,24] Within this approach the onsite coulomb term U, and the exchange term J, can always be grouped together into a single effective parameter (U-J),[23] and this effective parameter will simply be referred to as U in this paper. U values are obtained through self-consistent *ab initio* GGA calculations, using the methods described in Refs. [22,25,26]. The value U = 4.3 eV was used for Li$_x$FePO$_4$ and U = 4.5 eV for Li$_x$MnPO$_4$. These U values are the averages of U values obtained from the self-consistent calculations for the MPO$_4$ and LiMPO$_4$ states (U[LiFePO$_4$] = 3.71 eV, U[FePO$_4$] = 4.90 eV, U[LiMnPO$_4$] = 3.92 eV, U[MnPO$_4$] = 5.09 eV).[22]

## 3. Experimental Methods

*Sample preparation* - LiFePO$_4$ was prepared by solid-state reaction of Li$_2$CO$_3$ (99.999%, Alfa Aesar), FeC$_2$O$_4$·2H$_2$O (99.99%, Alfa Aesar) and NH$_4$H$_2$PO$_4$ (99.998%, Alfa Aesar). The appropriate amounts of these starting materials were ball-milled using zirconia milling media in acetone for a day, removed from the mill and dried, and ground in an argon-filled glovebox using an agate mortar and pestle. After recovering the mixture, it was first heated to 350°C for 10 hours in flowing argon. The calcined powder was then reground in an argon-filled glovebox and pressed into a pellet. The pellet was finally sintered at 700°C for 10 hours in Ar. X-ray diffraction was used to confirm the structure of the material and to make sure no impurity phases were present.

*UV-Vis-IR diffuse reflectance measurement* - Diffuse reflectance of the sintered polycrystalline LiFePO$_4$ pellet sample was measured over a wavelength range of 190–2000 nm (6.5eV-0.6eV), using a double-beam spectrophotometer (Model Cary 5E, Varian, Palo Alto, CA) with a diffuse reflectance accessory integrating sphere (Part #0010044900). Baseline spectra were collected using pressed polytetrafluoroethylene (PTFE) powder compacts (Product No. 04-101439-00, Varian) that were placed in the sample and reference beams. Data were collected at a scan rate of 600 nm/min with a data interval of 1.0 nm, a signal band width of 2.0 nm, and signal-averaging time of 0.1 s in UV-Vis range. In Near IR range, the data were collected at a scan rate of 2400nm/min with a data interval of 4.0 nm, a signal band width of 2.0 nm, and signal-averaging time of 0.1 s. Pellets were mounted in a blackened sample mask.

## 4. GGA and GGA+U Calculations of Electronic Structure

Figure 1 shows the FM total density of states (DOS) for $Li_xMPO_4$ (x = 0, 1 and M = Fe, Mn), with U = 0 (normal GGA) and U = 4.3 and 4.5, for Fe and Mn, respectively. Table 1 compares the band gaps in the different approximations, magnetic orderings, and materials, along with values from previous work. The calculated band gaps show some sensitivity to the choice of magnetic ordering, particularly for the $MnPO_4$ material, perhaps due to coupling of the magnetic ordering and the Jahn-Teller distortion (for more information on Jahn-Teller and magnetic coupling in $Mn^{3+}$ see Ref. [27]). However, it is clear that the qualitative impact of changing from LDA/GGA to DFT+U methods (here DFT+U is used to refer to both LDA+U and GGA+U methods) does not depend on the magnetic ordering. The previous work quoted in the last column of Table 1 is all obtained without U corrections, and therefore should be compared to the GGA data from this study. The results show that our pure GGA results are consistent with the ranges found in the previous literature. However, for precise comparison with the previous calculations the reader must consult the references provided, since different approximations, magnetic orderings, atomic positions, exchange correlation functions, *etc.* have been used. In general, more accurate treatment of the Coulombic correlations through GGA+U clearly yields larger band gaps, in some cases quite dramatically. This will be discussed further in Section 6.

## 5. Experimental Measurement of the Optical Band Gap of $LiFePO_4$

The diffuse reflectance spectrum for LiFePO$_4$ is shown in Figure 2. The large drop in reflectance below about 375 nm is due to absorption across the band gap of the material. Using the Kubelka-Munk remission function,[28] F=(1-R)$^2$/2R (where R is the diffuse reflectance), the gap can be determined as the energy at which F starts to increase linearly.[29] This gives 4.0 eV for the gap. An alternative way to determine the band gap is by extrapolating the onset of absorption to the wavelength axis (Shapiro's method), which gives a value of 3.8 eV.[30] We therefore take the measured band gap to be approximately 3.8-4.0 eV. This value is in very good agreement with the GGA+U result of 3.7 eV (see Table 1). The reflectance in the range beyond 600nm was also measured and no significant change in reflectance was detected up to 2000nm (corresponding to about 0.6 eV).

## 6. Discussion

The results in Figure 1 and Table 1 show that the electronic structure predicted by the GGA+U method can be quite different from that predicted by GGA. For example, the band gap is increased by more than a factor of 18 for LiFePO$_4$.

The DFT+U method was developed to help treat highly correlated electron systems, such as many transition metal oxides. The pioneering work of Anisimov, *et al.*[31] with DFT+U showed that the method generally gives much larger and more accurate band gaps for highly correlated transition metal oxides. More recent work has continued to support the power of DFT+U to improve on LDA/GGA results in an increasing number of highly correlated systems, including more accurate ordering of electronic states, band gaps, magnetic moments, atomic positions, and lattice parameters.[32-38] DFT+U methods generally improve over traditional LDA/GGA approaches when electrons are well localized. It is for this reason that the late transition metals, where the *d* orbitals are well localized, represent some of the most noted LDA/GGA failures.[32,39] It is likely that transition metal *d* orbitals are even more localized in the olivine phosphates than in simpler (non-phosphate) transition metal oxides. First, the transition metals themselves are very far apart, e.g., separated by 3.87 Å in LiFePO$_4$,[40] so direct hopping between them would be very minimal. Also, it is to be expected that the transition metal-*d* — oxygen-*p* hybridization in the phosphates is actually weaker than in simpler oxides. This is because phosphorous competes with the transition metal for bonding with the oxygen (this is the origin of the well known "inductive effect",[1,41] which increases the redox couple of the transition metal in the phosphates compared to the same metal in a simple oxide). Therefore, the weak interactions that encourage localized electrons in simple transition metal oxides will be even weaker in transition metal phosphates, possibly creating even greater errors in LDA/GGA calculations.

There is increasing direct evidence that some electronic structure properties of transition metal phosphates are not well reproduced by standard LDA/GGA calculations, and that DFT+U can greatly improve the accuracy of the calculations. For example, LDA/GGA predictions of the Li insertion voltage in both NASICON[42] and olivine[22,43] phosphate structures are quite far from experiment, and phase stability of lithiated phases in the Fe olivine phosphate is known to be in qualitative disagreement with experiment.[24] Both the phase stability and voltage problems have been shown to be corrected by using GGA+U methods.[22,24,43] In addition, incorrect predictions from LDA and GGA of the magnetic state of CoPO$_4$ were corrected using LDA+U and GGA+U[22,43].

The most definitive evidence that DFT+U is more accurate than LDA/GGA for band gaps in olivine phosphate materials comes from the combined experimental and computed results of this paper. For the case of LiFePO$_4$ the calculated band gap from GGA is somewhere in the range 0-0.3 eV, with some disagreement between different authors (see Table 1). However, the GGA+U result is about 3.7 eV, which is close to our measured value of 3.8-4.0 eV from diffuse reflectance spectroscopy. The combined evidence from bonding arguments, voltage, phase stability, magnetism, and band gap predictions, makes a clear case that DFT+U methods are necessary to obtain accurate results for many of the properties of this important class of materials.

Having established that the GGA+U method produces a more accurate DOS for this class of compounds, it is important to consider the implications of the DOS for the electronic conductivity. A large gap will lead to a very small number of intrinsically generated electrons or holes. For example, based on a 3.9 eV gap, LiFePO$_4$ would have an intrinsic electron (or hole) concentration of ~$10^{-14}$/cm$^3$ at room temperature (this result is calculated assuming parabolic bands and effective masses for electrons and holes that equal the bare electron mass, following Ref. [44]). Hence, the carrier concentration in this material will always be determined extrinsically, either by impurities, or more likely, by Li deficiency. Based on the volume of our LiFePO$_4$ unit cell, an under-stoichiometry of Li of only $10^{-35}$ Li per formula unit would create as many carriers as are intrinsically generated at 300 K. Since it is the electronic conductivity during the electrochemical cycling of Li in and out of the material that is relevant for battery application, the carrier density will always be determined by the amount of Li removed from the Li$_x$MPO$_4$ phases. *Therefore, as in most insulating intercalation materials, the band gap will not play any significant role in setting the concentration of conduction electrons or holes.*

Since the concentration of carriers is determined by the amount of Li, the key issue for conductivity will be the mobility of these carriers. Given the arguments above in support of localized electrons in this system, it is very likely that the key electrons involved in transport are not delocalized, but instead form localized small polarons. The relationship of conductivity to DOS for a polaronic conductor in the small-polaron model is fundamentally different than the relationship of conductivity to DOS in a material with delocalized electrons.[45-47] Polaron mobility is determined by the hopping rate of the polarons, which is a thermally activated process whose barrier is not directly related to the band gap.

A polaron conduction mechanism for $LiFePO_4$ is supported by the measured temperature dependence of the conductivity, which suggests that in the pure material electronic conductivity has an Arrhenius type behavior, with an activation energy of about 0.39-0.5 eV.[7,13] If this temperature dependence is due to exciting delocalized carriers over the gap, then based on the gap determined in this work the activation energy should be about 3.9/2 = 1.95 eV. Therefore, the measured activation energy is more likely a measure of the carrier mobility alone, with the carrier density set extrinsically, and independent of temperature.

In summary, our theoretical and experimental studies of the $Li_xFePO_4$ and $Li_xMnPO_4$ olivine materials reveal significant band gaps, largely induced by strong electron correlation at the transition metals. The structure and bonding of olivines, combined with the measured band gap and conduction activation energy in $LiFePO_4$, strongly suggest that *olivine conduction does not occur through the thermal creation of delocalized electrons across a small gap, but instead through a localized polaronic mechanism.* The number of carriers will be determined by the amount of Li off-stoichiometry that exists in the coexisting lithiated and delithiated phases during electrochemical Li cycling. Our results also indicate the dangers of using standard LDA/GGA theory to draw conclusions regarding the electronic structure and electronic conductivity of these materials. More accurate methods, such as DFT+U, previously shown to give more accurate predictions for a number of olivine properties,[22,24,43] are required for these highly correlated systems.

## Acknowledgments

We gratefully acknowledge support from the Department of Energy under grant DE-FG02-96ER45571, the Assistant Secretary for Energy Efficiency and Renewable Energy, Office of FreedomCAR and Vehicle Technologies of the U.S. Department of Energy via subcontract number 6517748, the National Science Foundation (MRSEC Program) under contract DMR-0213282, and computing resources from the National Partnership for Advanced Computational Infrastructure (NPACI).

40 A. S. Andersson, B. Kalska, L. Haggstrom, and J. O. Thomas, Solid State Ionics **130**, 41 (2000).
41 A. K. Padhi, K. S. Nanjundaswamy, C. Masquelier, and J. B. Goodenough, J. Electrochem. Soc. **144**, 2581 (1997).
42 D. Morgan, G. Ceder, M. Y. Saidi, J. Barker, J. Swoyer, H. Huang, and G. Adamson, Chem. Mater. **14**, 4684 (2002).
43 O. L. Bacq, A. Pasturel, and O. Bengone, Phys. Rev. B **69**, 245107 (2004).
44 N. W. Aschcroft and N. D. Mermin, *Solid State Physics* (Harcourt Brace, New York, 1976).
45 W. D. Kingery, H. K. Bowen, and D. R. Uhlman, *Introduction to Ceramics* (John Wiley & Sons, New York, 1976).
46 S. R. Elliott, *Physics of amorphous materials* (Longman, London, 1983).
47 I. G. Austin and N. F. Mott, Advances in Physics **50**, 757 (2001).
**13**

| System | GGA Gap (FM) | GGA+U Gap (FM) | GGA Gap (AFM) | GGA+U Gap (AFM) | Other work (FM) |
|---|---|---|---|---|---|
| FePO4 | 0.5 | 1.9 | 0.9 | 2.1 | 0.4 [6], 0.3 (0.7 AFM)[11] |
| LiFePO4 | 0.2 | 3.7 | 0.2 | 3.7 | 0.3[6], 0.0 (0.0 AFM)[11], 0.2[12], 0.0[13] |
| MnPO4 | 0.1 | 0.5 | 0.9 | 1.1 | — |
| LiMnPO4 | 2.0 | 3.8 | 2.1 | 4.0 | 2[6], 1.7[11] |

Table 1: Calculated band gaps for Fe and Mn olivine, from this work and previous references. All energies are in eV. All results from other work are ferromagnetic (FM) unless specified as antiferromagnetic (AFM), and GGA results are quoted when available. Previous results are obtained with a range of different methods and approximations, and therefore the original references must be consulted for precise comparisons with each other or the present work. However, the previous results are useful for establishing the range of values obtained using traditional LDA and GGA methods.

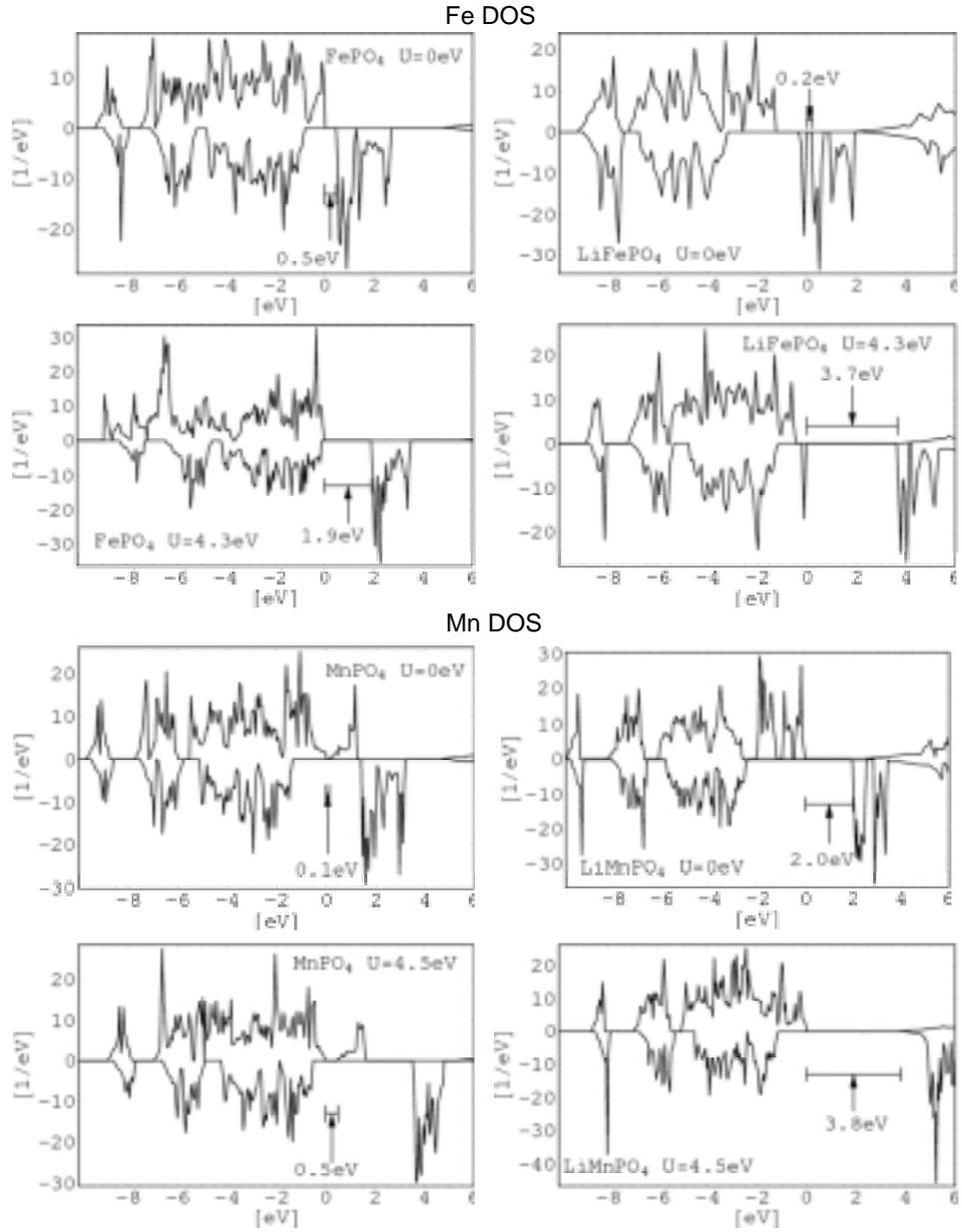

Figure 1: The ferromagnetic total density of states for $Li_xMPO_4$ (x = 0, 1 and M = Fe, Mn), with U = 0 (normal GGA) and U = 4.3 and 4.5, for Fe and Mn, respectively. The positive (negative) axis is the majority (minority) spin direction. Note that for Mn $PO_4$ the FM

gaps are considerably different for AFM gaps in Table 1.

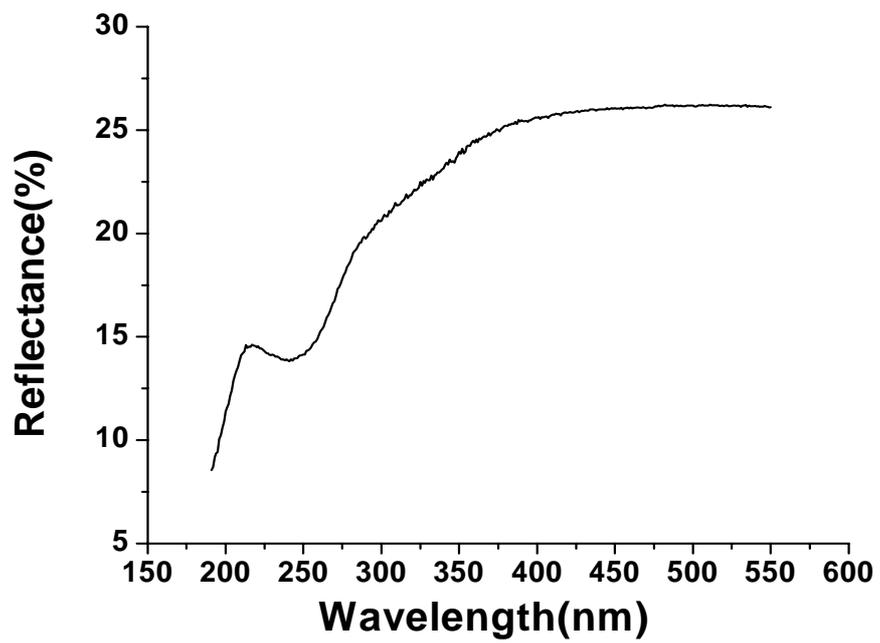

Figure 2: The diffuse reflectance spectrum for LiFePO$_4$.